%% file: main.tex
\documentclass[a4paper,
biblatex,     
keeplastbox,   
]{jacow}
\usepackage{pdfpages,multirow,ragged2e} %
\usepackage{tikz,algorithm,algpseudocode}
\usepackage{hyperref}
\makeatletter
\renewcommand{\ALG@name}{Specification}
\makeatother
\usepackage{listings,xcolor,graphicx,nameref}
\definecolor{dkgreen}{rgb}{0,0.6,0}
\definecolor{gray}{rgb}{0.5,0.5,0.5}
\definecolor{mauve}{rgb}{0.58,0,0.82}
\definecolor{black}{rgb}{0,0,0}

\lstdefinestyle{XML}{
    language=XML, morekeywords={Name, Type, UId},
     basicstyle=\ttfamily\scriptsize, breaklines=true, frame=single,keywordstyle=\color{blue},
	commentstyle=\color{dkgreen},
	stringstyle=\color{mauve}
}

\lstdefinestyle{assertion}{
    language=java,
     basicstyle=\ttfamily\scriptsize, breaklines=true, frame=single,
	commentstyle=\color{black}
}

\makeatletter%
\ifboolexpr{bool{xetex}}
{\renewcommand{\Gin@extensions}{.pdf,%
		.png,.jpg,.bmp,.pict,.tif,.psd,.mac,.sga,.tga,.gif,%
		.eps,.ps,%
}}{}
\makeatother

\ifboolexpr{bool{xetex} or bool{luatex}} 
{}                                      
{\usepackage[utf8]{inputenc}}           

\usepackage[USenglish]{babel}

\ifboolexpr{bool{jacowbiblatex}}%
{%
	\addbibresource{Bibliography.bib}
}{}
\newlist{inlinelist}{enumerate*}{1}
\setlist*[inlinelist,1]{%
	label=\textit{(\roman*)},
}

\newlist{inlinelistbullet}{enumerate*}{1}
\setlist*[inlinelistbullet,1]{%
	label=$\bullet$,
}
\usepackage{ccicons}
\usepackage{eso-pic}
\usepackage{graphicx}
\definecolor{colorMarge}{RGB}{255,255,255}
\newlength{\distance}
\newlength{\rulethickness}
\newlength{\ruleheight}
\newlength{\xoffset}
\newlength{\yoffset}
\AddToShipoutPicture{%
  \setlength\fboxsep{0pt}
  \ifodd\value{page}%
    \setlength{\xoffset}{\dimexpr\paperwidth-\rulethickness-\distance}%
  \else
    \setlength{\xoffset}{\distance}%
  \fi
  \AtPageLowerLeft{%
    \put(\LenToUnit{\xoffset},\LenToUnit{\yoffset}){%
      \colorbox{colorMarge}{%
        \parbox[b][\ruleheight][b]{\rulethickness}{%
          \centering

            \rotatebox[origin=lB]{90}{\small{\strut \quad\quad\quad \ccby \quad Content from this work may be used under the terms of the CC BY 3.0 licence (© 2022). Any distribution of this work must maintain attribution to the author(s), title of the work, publisher, and DOI}}%

        }%
      }%
    }%
  }%
}

\usepackage[absolute]{textpos}
\begin{document}
	
	\begin{textblock}{10}(2,2)
\noindent\small 18th Int. Conf. on Acc. and Large Exp. Physics Control Systems\\
ISBN: 978-3-95450-221-9 \quad\quad\quad\quad\quad\quad\quad\quad ISSN: 2226-0358
\end{textblock}
\begin{textblock}{10}(11.4,2)
\noindent\small ICALEPCS2021, Shanghai, China \quad\quad JACoW Publishing\\
doi:10.18429/JACoW-ICALEPCS2021-WEPV042
\end{textblock}
	
	\title{Applying model checking to highly-configurable safety critical software: the SPS-PPS PLC program}
	
	\author{B. Fern\'andez\thanks{borja.fernandez.adiego@cern.ch}, I. D. Lopez-Miguel, J-C. Tournier, E. Blanco, \\
		T. Ladzinski, F. Havart, CERN, Geneva, Switzerland}
	
	\maketitle

	\begin{abstract}
		An important aspect of many particle accelerators is the constant evolution and frequent configuration changes that are needed to perform the experiments they are designed for.
		
		This often leads to the design of configurable software that can absorb these changes and perform the required control and protection actions. This design strategy minimizes the engineering and maintenance costs, but it makes the software verification activities more challenging since safety properties must be guaranteed for any of the possible configurations.
		
		Software model checking is a popular automated verification technique in many industries. This verification method explores all possible combinations of the system model to guarantee its compliance with certain properties or specification. This is a very appropriate technique for highly configurable software, since there is usually an enormous amount of combinations to be checked.
		
		This paper presents how PLCverif, a CERN model checking platform, has been applied to a highly configurable Programmable Logic Controller (PLC) program, the SPS Personnel Protection System (PPS). The benefits and challenges of this verification approach are also discussed.
	\end{abstract}

	\section{Introduction}
	
	Model checking is a popular verification technique that has been applied in many industries to guarantee that a critical system meets the specifications. 
	Model checking algorithms explore all possible combinations of a system model, trying to find a violation of the formalized specification in the model. This technique is very appropriate for highly configurable projects, since it is necessary to guarantee the safety of the system for all possible configurations.
	
	When model checking shows a discrepancy between the PLC program and the specification, it means that either the PLC program has a bug or the specification is incomplete or incorrect.
	
	In the domain of critical PLC programs, several researchers and engineers published their experiences in the field. To name but a few, in \cite{Pavlovic:ICST2010} the authors translate the PLC program of an interlocking railway system, written in the FBD (Function Block Diagrams) language, into the input format language of NuSMV to verify their specification written as CTL (Computation Tree Logic) properties. In \cite{Moreira:MSc}, the PLC program that controls the doors’ opening and closing in the trains from the Metro in Brasília, Brazil, was formally verified. In this case, a B model \cite{Abrial:Bbook} is created automatically from the PLC code. This model is formally verified using the model checker ProB \cite{Leuschel:FME03}.
	
	When applying model checking to PLC programs, three main challenges are faced: (1) building the mathematical model of the program, (2) formalizing the requirements to be checked and (3) the state-space explosion, i.e. the number of possible input combinations and execution traces is too big to be exhaustively explored. In our case, we use the open-source tool PLCverif\cite{BlancoViñuela:2777799}, developed at CERN. It creates automatically the models out of the PLC program and integrates three state-of-the-art model checkers: nuXmv \cite{Cavada:CAV2014}, Theta \cite{Toth:FMCAD2017} and CBMC \cite{Clarke:TACAS2004}. It also implements some reduction and abstraction mechanisms to reduce the number of states to be explored and to speed up the verification. Therefore, challenges 1 and 3 are transparent for the user. There are certainly still limitations and large state-space PLC program models cannot be verified. Regarding challenge 2, PLCverif also provides mechanisms to help the users to formalize their requirements and provide a precise specification. However, this is normally a difficult task, specially for configurable programs. In this paper, we will show examples of functional requirements formalization with PLCverif.
	More details about PLCverif can be found in \cite{Lopez:ICALEPCS2021,BlancoViñuela:2777799}.
	
	This paper aims to show the benefits of applying model checking to verify highly configurable PLC programs. In such systems, it is usually unfeasible to check all possible combinations by traditional testing methods and model checking is a good complement to these methods, especially for module verification.
	
	In particular, this paper shows how PLCverif was applied to the PLC programs of the SPS Personnel Protection System (PPS) \cite{Ladzinski:ICALEPCS2019} and how it helped to improve their original specification and correct PLC bugs before the commissioning of the system.

	

	\section{SPS Personnel Protection System} \label{sec:SPS}
	
	The SPS-PPS is a large distributed control system in charge of the access control and the personnel protection of the SPS accelerator. 
	
	The SPS has 16 access zones divided in different sectors and each access zone has an access point. Several access zones are always interlocked with the same elements inhibiting operation with beam when a hazardous event is detected. This is the concept of a safety chain. Each safety chain contains the "important safety elements" to stop the beam (EISb) when a hazardous event is detected or to avoid access (EISa) when the accelerator is in Beam mode.
	
	All details of the system can be found in \cite{Ladzinski:ICALEPCS2019}.
	
	The SPS evolves with time as new access zones are added or existing ones modified. For that reason, a configurable system was designed.
	
	\subsection{Hardware architecture}
	
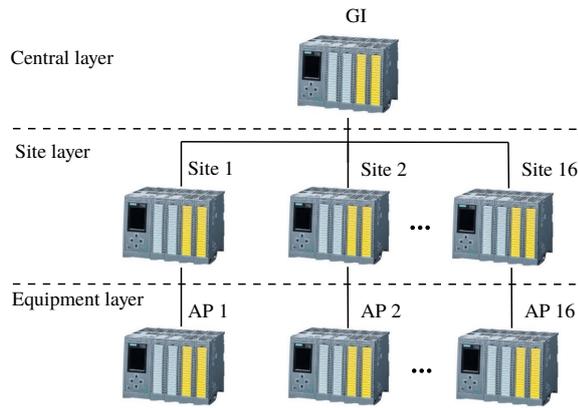
\begin{figure}
		\centering
		\scalebox{.75}{\input{WEPV042f1.tex}}
		\caption{Simplified SPS-PPS PLC interlock architecture}
		\label{fig:architecture}
	\end{figure}

	The SPS-PPS safety-interlock part is based on Siemens S7-1500F series of PLC controllers. It is formed by the following 3 layers as shown in Figure \ref{fig:architecture}:
	\begin{Itemize}
		\item The central layer is composed by the Global Interlock (GI) PLC. This PLC receives the information of all the Site PLCs and computes, for example, the conditions to allow Beam or Access modes to each safety chain and the conditions to apply veto to all the EISa and EISb associated to the safety chain. 
		\item  The site layer is composed by 16 Site PLCs (1 PLC per access zone). These PLCs monitor the EISa devices (doors, emergency handles, etc.), receive the status from the AP PLCs, perform some automatic control procedures (e.g. patrol management) and apply the safety interlock actions (e.g. stop the beam) with the commands received from the GI PLC for each access zone.
		\item  The equipment layer is composed by 16 Access Point (AP) PLCs (1 PLC per access point). They monitor and control the PAD (Personnel Access Device) and the MAD (Material Access Device) of the access point.
	\end{Itemize}
	
	\subsection{Software architecture}
	All Site PLCs run the same generic PLC program, a common logic for all of them with a specific configuration for each access zone.
	The logic is composed by a set of Functions (FC) and Function Blocks (FBs) executed sequentially from a cyclic interrupt every 300 ms.
	Each access zone has a different configuration, indicating for example how many EISa and EISb are installed in each of them, in which sector they are located and to which safety chain they belong.
	
	A similar approach is in place for the AP PLCs.
	
	The GI PLC contains less configurable parameters since it communicates with all Site PLCs, having a global view of the system.	
	
	\subsection{Software modules specification}
	The different program modules (FC and FBs) are specified in a detailed design document. It contains 106 formulas that describe the requirements for the 29 FC and FBs.
	
	The chosen formalism for the specification is a simple \textit{if-else} conditional statement that contains Boolean formulas, as shown in the example specification \ref{alg:specGeneric}. The goal was to have a simple pseudo-formal specification providing precise and unambiguous requirements for each module of the system.
	
	\begin{algorithm}
	\scriptsize
		\caption{Template for specifying protection actions of the Safety program}\label{alg:specGeneric}
		\begin{algorithmic}
			
			\If {(\textit{Boolean expression condition})}
			\State $result \gets 0$ 
			\ElsIf {(\textit{Boolean expression condition 2})}
			\State $result \gets 1$ 
			\Else
			\State $result \gets result$ 
			\EndIf
			
		\end{algorithmic}
	\end{algorithm}
	
	
	\section{Modules verification with PLCverif} \label{sec:PLCverif}
	PLCverif was the tool to apply model checking to the SPS-PPS modules (FC and FBs).
	All the necessary model transformations to create the formal model of the PLC program are done automatically. The user only needs to provide the exported PLC program from the programming environment tool, Siemens TIA Portal\footnote{\url{https://new.siemens.com/global/en/products/automation/industry-software/automation-software/tia-portal/software.html}}, import it into PLCverif and finally formalize the requirements to be verified.
	
	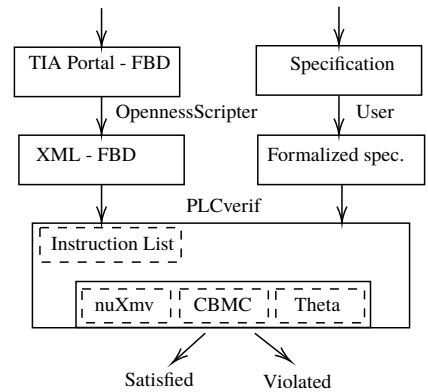
\begin{figure}
		\centering
		\scalebox{.75}{\input{WEPV042f2.tex}}
		\caption{PLCverif workflow to verify TIA portal Safety programs}
		\label{fig:Flow}
	\end{figure}
	
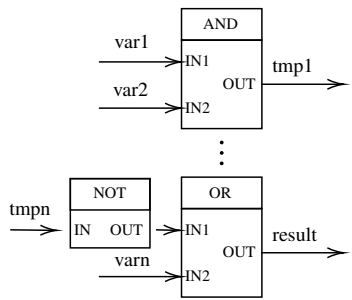
\begin{figure}
		\centering
		\scalebox{.75}{\input{WEPV042f3.tex}}
		\caption{FBD program example}
		\label{fig:SIF2-1-FBD}
	\end{figure}

	Figure \ref{fig:Flow} shows the workflow to verify a FC or FB from a TIA Portal safety PLC project with PLCverif:
	\begin{enumerate}
		\item The project is hosted in TIA Portal and the PLC program is written in the FBD (Function Block Diagram) programming language. An example of how this language looks like is shown in Figure \ref{fig:SIF2-1-FBD}. Here an AND gate and an OR gate are displayed.
		\item Taking advantage of the Siemens OpennessScripter tool\footnote{\url{https://support.industry.siemens.com/cs/us/en/view/109742322}}, the program is exported to XML (\textit{F\_FBD} XML format). An example of how an AND gate is represented in this language is shown in listing \ref{listing:XML}.
		\item PLCverif can then import that XML code, translating it into STL code, which can be understood by PLCverif. Subsequently, it creates the formal models.
		\item From the specification created by the process engineers, the user needs to formalize the requirements, such as the assertion shown in Listing \ref{eq:assertion}. This can be direct input to PLCverif.
		\item Finally, PLCverif executes one of the model checking backends and generates a report, stating whether the property is satisfied or violated. In case it is violated, it shows the trace leading to that property failure (the counterexample).
	\end{enumerate}
	
	\begin{lstlisting}[caption={XML code exported from TIA portal representing an AND gate}, style=XML, label=listing:XML]
<!-- Input variables -->
<Access Scope="LocalVariable" UId="21">
<Symbol>
<Component Name="var1"/>
</Symbol>
</Access>
<Access Scope="LocalVariable" UId="22">
<Symbol>
<Component Name="var2"/>
</Symbol>
</Access>

<!-- AND block -->
<Part Name="And" UId="97">
<TemplateValue Name="Card" Type="Cardinality">2</TemplateValue>
<TemplateValue Name="SrcType" Type="Type">Word</TemplateValue>
</Part>

<!-- Connecting the inputs with the block -->
<Wire UId="134">
<IdentCon UId="21" />
<NameCon UId="97" Name="in1" />
</Wire>
<Wire UId="135">
<IdentCon UId="22" />
<NameCon UId="97" Name="in2" />
</Wire>

<!-- Output variable -->
<Wire UId="136">
<NameCon UId="97" Name="out" />
<IdentCon UId="23" />
</Wire>
	
	\end{lstlisting}

	\begin{lstlisting}[caption={Example of a PLCverif assertion}, style=assertion, label=eq:assertion]
//#ASSERT (var1 AND var2) = result;
	\end{lstlisting}
	
	\section{Verification results and analysis} \label{sec:results}
	
	Although various safety functions were verified during this project, due to space limitations for this paper, only two of them will be shown here: the \textit{SIF-X1} Function from the Site PLCs and the \textit{SIF-2} Function from the GI PLC. Each of these PLC Functions implements several requirements of a similar nature and each of these requirements was verified with PLCverif by including an assertion which represents the mathematical Boolean formula given for that requirement. 
	After executing a verification case with PLCverif, the following outcomes can occur:
	
	\begin{itemize}
		\item Satisfied property. It will be shown in \hyperref[sec:example1]{example 1} from SIF-X1.
		\item Violated property. The reasons why this happens need to be further investigated, leading to the following two possibilities:
		\begin{itemize}
			\item Incomplete specification. \hyperref[sec:example2]{Example 2} from SIF-2.
			\item Bug in the program. \hyperref[sec:example3]{Example 3} from SIF-2.
		\end{itemize}
	\end{itemize}

	\subsection{SIF-X1 Function}
	
    This function is in charge of monitoring all the EISa of the access zone and computing the status (safe or unsafe) for each safety chain.

	\subsubsection{Example 1 - Satisfied property.} \label{sec:example1}
	
	The selected requirement for this use case is the following: 
	For each Safety Chain, this Function monitors all the EISa that are installed in the access zone and belong to the Safety Chain. It also receives the status from its access point (PAD, MAD and token distributor status). If all these elements are in a safe state, it returns a Boolean variable indicating that the Safety Chain is safe. 
	
	The formalized requirement is shown in specification \ref{alg:SIFX1}, where 
	\vspace{-0.1cm}
	\begin{description}
	\itemsep0em 
	    \item $i$ is the EISa index,
	    \item $j$ is the safety chain index,
	    \item $N_i$ is the number of EISa assigned to the safety chain $j$ (the maximum number of EISa per access zone is 32),
	    \item $I\_EISa\_Pos[i]\_Stat$ represents the status of the EISa $i$: true if $i$ is closed,
	    \item $I\_EISa\_PU[i]\_Stat$ represents the status of the emergency handle of the EISa $i$: true if the emergency handle is armed, $I\_EISa\_Pos[i]$ is a configuration variable: true if the EISa $i$ is installed in the access zone,
	    \item $I\_EISa\_PU[i]$ is configuration variable: true if the EISa $i$ has an emergency handle,
	    \item $SC-S{j}[i]$ is configuration variable: true if the EISa $i$ belongs to the Safety Chain $j$,
	    \item $S0\_AP\_Pos[j]$, $S0\_AP\_PU[j]$ and $S0\_AP\_Key\_Distrib[j]$ represent the status of the access point: true if the access point is closed, the emergency handle is armed and the token distributor is safe respectively,
	    \item $N\_EISa\_Safe[j]$ is the output variable of the Function and represents the status of the safety chain $j$: true if $j$ is safe.
	\end{description}
	
	\begin{algorithm}
	\scriptsize
		\caption{Partial, simplified requirement of SIF-X1}\label{alg:SIFX1}
		\begin{algorithmic}
			
			\If {
				\big(
				\\
				$
				\displaystyle 
				\prod_{i = 1}^{N_i} \displaystyle \big(I\_EISa\_Pos\_Stat[i]\ \vee\ I\_EISa\_Bypass[i]\big)
				$\\
				\noindent
				$
				\scriptstyle \forall i \in [1,N_i] : (I\_EISa\_Pos[i] = 1\  \wedge\ SC-S\{j\}[i] = 1
				$\\
				\noindent
				$
				\displaystyle \wedge
				$\\
				\noindent
				$
				\displaystyle 
				\prod_{i = 1}^{N_i} \displaystyle \big(I\_EISa\_PU\_Stat[i]\ \vee\ I\_EISa\_Bypass[i]\big)
				$\\
				$
				\scriptstyle \forall i \in [1,N_i] : (I\_EISa\_PU[i] = 1\  \wedge\ SC-S\{j\}[i] = 1
				$\\
				\noindent
				$
				\displaystyle \wedge
				$\\
				\noindent
				$
				\displaystyle S0\_AP\_Pos[j] 
				$     
				\noindent
				$
				\displaystyle \wedge
				$
				\noindent
				$
				\displaystyle S0\_AP\_PU[j]
				$     
				\noindent
				$
				\displaystyle \wedge
				$
				\noindent
				$
			    	\displaystyle S0\_AP\_Key\_Distrib[j]
				$    
				\\
				\big)
			}
			\State $\displaystyle N\_EISa\_Safe[j]  \gets 1$
			\Else
			\State $\displaystyle N\_EISa\_Safe[j]  \gets 0$
			\EndIf
			
		\end{algorithmic}
	\end{algorithm}
	
	The behaviour of this relatively complex specification is shown in Table \ref{tab:SIFX1} with an example. Here, 2 doors (EISa) are installed in this access zone (indexes 1 and 4 of $EISa\_Pos[i]$), only one of them has an emergency handle (index 1 of $EISa\_PU[i]$). One of the EISa belongs to the Safety Chain 0 (index 1 of $SC-S\_0[i]$) and the other one to the Safety Chain 1 (index 4 of $SC-S\_1[i]$). For each safety chain $j$, the resultant $N\_EISa\_Safe[j]$ is true if the installed doors and emergency handles are in the safe state ($EISa\_Pos\_Stat[i]$ and $EISa\_PU\_Stat[i]$).
	
	\begin{table}[!hbt]
	\scriptsize
		\centering
		\caption{SIF-X1 expected behaviour}\label{tab:SIFX1}
		\begin{tabular}{lcc}
			\toprule
			\textbf{Source} & \textbf{Variable} & \textbf{Value} \\
			\midrule
			Input   & $EISa\_Pos\_Stat$ & 0000 0000 0000 1001        \\ 
			Input & $EISa\_PU\_Stat$  & 0000 0000 0000 0001        \\ 
			... & ...             & ...        \\ 
			Input (AP PLC) & $S0\_AP\_Pos$     & 0000 0000 0000 1001        \\ 
			... & ...             & ...        \\ 
			Configuration    & $EISa\_Pos$       & 0000 0000 0000 1001        \\ 
			Configuration & $EISa\_PU$        & 0000 0000 0000 0001        \\ 
			... & ...             & ...        \\ 
			Configuration     & $SC-S\_0$         & 0000 0000 0000 0001        \\ 
			Configuration & $SC-S\_1$         & 0000 0000 0000 1000        \\ 
			... & ...             & ...        \\ 
			Output & $N\_EISa\_Safe$   & 0000 0000 0000 1001        \\ 
			\bottomrule
		\end{tabular}
		\label{tab:margins}
	\end{table}


	The specification was translated into 16 verification cases (one per safety chain), in order to guarantee that this property is respected in all safety chains for all possible combinations of the input and configuration variables. The example for the safety chain 0 can be seen in listing \ref{eq:assertion1} (simplified assertion).
	
	The corresponding PLC program has 94 WORD (16-bit variable) and 4 BOOL configuration variables, and 21 WORD and 2 BOOL input variables to implement all the requirements. This implies approximately $2^{16 \cdot (94+21)+6} = 2^{1846} \approx 5.0 \cdot 10^{555}$ combinations to be checked.
	The equivalent STL code of the original F-FBD version is around 700 lines of code.
	
	PLCverif was able to validate the 16 verification cases in 41 seconds. The result of all properties (assertions) was satisfied. This means that no counterexamples were found and, hence, the properties always hold true.

	\begin{lstlisting}[caption={SIF-X1 formalized requirement for PLCverif.}, style=assertion, label=eq:assertion1]
//#ASSERT
(((I_EISa_Pos_Stat OR I_EISa_Bypass) AND
(SC-S_0 AND I_EISa_Pos))
= (SC-S_0 OR I_EISa_Pos)) AND
(((I_EISa_PU_Stat OR I_EISa_Bypass) AND
(SC-S_0 AND I_EISa_PU))
= (SC-S_0 OR I_EISa_PU)) AND
(S0_AP_Key_Distrib.%X0 AND
S0_AP_PU.%X0 AND S0_AP_Pos.%X0)
= N_EISa_Safe.%X0
	\end{lstlisting}

	\subsection{SIF-2 Function}
	
	The following two examples belong to the verification of the SIF-2 Function. In this function, PLCverif was able to find several discrepancies between the PLC program and the specification and two of them are presented here.
	
	This function contains the transitions rules between the Beam and Access modes for each safety chain $j$. In order to allow the transition between modes, the function checks the status of the safety chain (computed variable calculated by another GI PLC function) and the requests from the operator, via the OKP (Operator Key Panel). When all safety conditions are met, this function authorizes the change of mode. 
	The function also contains the conditions to lock or release the keys from the OKP.
	
	The implementation of the SIF-2 Function Block has 19 WORD and 1 BOOL input variables to implement all the requirements. This implies approximately $2^{19\cdot 16 + 1} = 2^{305} \approx 6.5 \cdot 10^{91}$ combinations to be checked.
	
	The equivalent STL code of the original F-FBD version is around 1200 lines of code.
	
	\subsubsection{Example 2 - Incomplete specification.}\label{sec:example2}

	This specification was related to the conditions to switch to Beam mode. The property was formalized with the assertion shown in listing \ref{eq:assertion2}:
	
	\begin{lstlisting}[caption={SIF-2 Beam mode activation formalized requirement for PLCverif}, style=assertion, label=eq:assertion2]
//#ASSERT
((NOT((N_EXT_ACCE_OK AND N_NO-SAFETY_ERR) AND
((I_PB_TEST_ON AND I_Key_TEST) OR
(I_PB_Acce_ON AND I_Key_Acce)))
OR (NOT N_MODE_BEAM)) = 16#FFFF)
	\end{lstlisting}

	This assertion was verified with PLCverif in 93 seconds and a counterexample was found, meaning that the property was violated. 
	The reason for this violation was a missing variable in the formula from the documentation.
	
	\subsubsection{Example 3 - Bug in the PLC program.}\label{sec:example3}
	
	
	In this case, the conditions to release an access key from the OKP for one of the safety chains were used as specification. The formalized assertion was the one shown in listing \ref{eq:assertion3}:
	
	\begin{lstlisting}[caption={SIF-2 extractors access key release formalized requirement for PLCverif}, style=assertion, label=eq:assertion3]
//#ASSERT
((N_SECU_NO-REQ_Down.%X15 AND N_EXT_BEAM_OK.%X15)
= (SIF-2_TAGS.O_RLS_ACCESS.%X15))
	\end{lstlisting}
	
	PLCverif was able to verify this assertion in 147 seconds. The result was not satisfied and a counterexample was provided. After analysing the counterexample and discussing with the SPS-PPS experts, it was concluded that there was a problem in the PLC program. The program was modified accordingly to include the missing variables before the commissioning of the system.
	

	
	\subsection{Analysis}
	
	PLCverif hides most of the complexity of applying model checking to PLC programs. Nevertheless, a few manual steps were still needed in this project. In particular to import the program in PLCverif and to create the formal properties from the pseudo-formalized specification. 
		
	We also encounter the state-space explosion problem in other functions, where the number of configurations and input variables was much bigger. 
	
	Overall, PLCverif helped to validate a few critical functions from the SPS-PPS PLC program and it was a good complement to the testing activities performed in the lab by the access control experts at CERN (e.g. the SIF-X1 function).
	
	PLCverif was able to find discrepancies between some of the specifications and the PLC program before the commissioning of the system, sometimes due to an incomplete or erroneous specification, other times due to a bug in the program (e.g. the SIF-2 function).
	Due to the complexity of some of the specifications, it was sometimes very hard for experts to validate the desired behaviour. By using model checking, the experts were able to find undesirable corner cases, helping them to understand the behaviour of the system and modifying the specification when necessary.

	\section{Conclusions} \label{sec:conclusions}
	This paper presented a highly configurable PLC program and the benefits and challenges of applying model checking to verify it.
	In general, we can conclude that model checking contributed to this project to (1) detect bugs in the PLC program before the commissioning, (2) identify deficiencies in the specification, and (3) help experts to better understand the behaviour of the program for all possible configurations.
	
	
	There are still several challenges to overcome for model checking to become a common practice in the industrial automation domain. Probably the more important ones are: (1) the integration of model checking in PLC programming environments (e.g. TIA portal) to minimize the number of manual steps and (2) the improvement of verification performance with better algorithms and abstraction techniques.
	
	In the context of the PLCverif project, the future work will focus on the following directions: (1) improve PLCverif to reduce the verification time, (2) support more PLC manufacturers -- we are currently working on supporting Schneider ST and IL programs. More information about the current and future work of PLCverif can be found in \cite{Lopez:ICALEPCS2021}.


	
	%
	%
	\ifboolexpr{bool{jacowbiblatex}}%
	{\printbibliography}%
	{%
	} 
	%
	%
	
	
\end{document}

%% file: WEPV042f1.tex
\tikzset{every picture/.style={line width=0.75pt}} 

\begin{tikzpicture}[x=0.75pt,y=0.75pt,yscale=-1,xscale=1]

\draw (225,49) node  {\includegraphics[width=52.5pt,height=52.5pt]{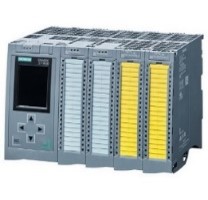}};
\draw (109,151) node  {\includegraphics[width=52.5pt,height=52.5pt]{WEPV042f4.jpg}};
\draw (219,151) node  {\includegraphics[width=52.5pt,height=52.5pt]{WEPV042f4.jpg}};
\draw (328,152) node  {\includegraphics[width=52.5pt,height=52.5pt]{WEPV042f4.jpg}};
\draw    (225,108.57) -- (225,78.57) ;
\draw    (331.71,95.04) -- (114,94.57) ;
\draw    (114,124.57) -- (114,94.57) ;
\draw    (225,125.57) -- (225,95.57) ;
\draw    (332,125.79) -- (331.71,95.04) ;
\draw  [dash pattern={on 3.75pt off 3.75pt}]  (2,85.79) -- (377.43,86.21) ;
\draw  [dash pattern={on 3.75pt off 3.75pt}]  (1.9,190) -- (376.43,190.21) ;
\draw (109,245) node  {\includegraphics[width=52.5pt,height=52.5pt]{WEPV042f4.jpg}};
\draw (219,245) node  {\includegraphics[width=52.5pt,height=52.5pt]{WEPV042f4.jpg}};
\draw (331,246) node  {\includegraphics[width=52.5pt,height=52.5pt]{WEPV042f4.jpg}};
\draw    (114,218.57) -- (114,178.93) ;
\draw    (225,219.57) -- (224.8,178.49) ;
\draw    (334,219.57) -- (333.83,179.09) ;
\draw  [fill={rgb, 255:red, 0; green, 0; blue, 0 }  ,fill opacity=1 ] (268,248.21) .. controls (268,247.66) and (268.45,247.21) .. (269,247.21) .. controls (269.55,247.21) and (270,247.66) .. (270,248.21) .. controls (270,248.77) and (269.55,249.21) .. (269,249.21) .. controls (268.45,249.21) and (268,248.77) .. (268,248.21) -- cycle ;
\draw  [fill={rgb, 255:red, 0; green, 0; blue, 0 }  ,fill opacity=1 ] (278,248.21) .. controls (278,247.66) and (278.45,247.21) .. (279,247.21) .. controls (279.55,247.21) and (280,247.66) .. (280,248.21) .. controls (280,248.77) and (279.55,249.21) .. (279,249.21) .. controls (278.45,249.21) and (278,248.77) .. (278,248.21) -- cycle ;
\draw  [fill={rgb, 255:red, 0; green, 0; blue, 0 }  ,fill opacity=1 ] (273,248.21) .. controls (273,247.66) and (273.45,247.21) .. (274,247.21) .. controls (274.55,247.21) and (275,247.66) .. (275,248.21) .. controls (275,248.77) and (274.55,249.21) .. (274,249.21) .. controls (273.45,249.21) and (273,248.77) .. (273,248.21) -- cycle ;
\draw  [fill={rgb, 255:red, 0; green, 0; blue, 0 }  ,fill opacity=1 ] (267,152.21) .. controls (267,151.66) and (267.45,151.21) .. (268,151.21) .. controls (268.55,151.21) and (269,151.66) .. (269,152.21) .. controls (269,152.77) and (268.55,153.21) .. (268,153.21) .. controls (267.45,153.21) and (267,152.77) .. (267,152.21) -- cycle ;
\draw  [fill={rgb, 255:red, 0; green, 0; blue, 0 }  ,fill opacity=1 ] (277,152.21) .. controls (277,151.66) and (277.45,151.21) .. (278,151.21) .. controls (278.55,151.21) and (279,151.66) .. (279,152.21) .. controls (279,152.77) and (278.55,153.21) .. (278,153.21) .. controls (277.45,153.21) and (277,152.77) .. (277,152.21) -- cycle ;
\draw  [fill={rgb, 255:red, 0; green, 0; blue, 0 }  ,fill opacity=1 ] (272,152.21) .. controls (272,151.66) and (272.45,151.21) .. (273,151.21) .. controls (273.55,151.21) and (274,151.66) .. (274,152.21) .. controls (274,152.77) and (273.55,153.21) .. (273,153.21) .. controls (272.45,153.21) and (272,152.77) .. (272,152.21) -- cycle ;

\draw (117,106) node [anchor=north west][inner sep=0.75pt]   [align=left] {Site 1};
\draw (232,107) node [anchor=north west][inner sep=0.75pt]   [align=left] {Site 2};
\draw (339,107) node [anchor=north west][inner sep=0.75pt]   [align=left] {Site 16};
\draw (-1,32) node [anchor=north west][inner sep=0.75pt]   [align=left] {Central layer};
\draw (2,94.79) node [anchor=north west][inner sep=0.75pt]   [align=left] {Site layer};
\draw (222,4) node [anchor=north west][inner sep=0.75pt]   [align=left] {GI};
\draw (0,194) node [anchor=north west][inner sep=0.75pt]   [align=left] {Equipment layer};
\draw (117,202) node [anchor=north west][inner sep=0.75pt]   [align=left] {AP 1};
\draw (232,202) node [anchor=north west][inner sep=0.75pt]   [align=left] {AP 2};
\draw (341,202) node [anchor=north west][inner sep=0.75pt]   [align=left] {AP 16};

\end{tikzpicture}

%% file: WEPV042f2.tex
\tikzset{every picture/.style={line width=0.75pt}} 

\begin{tikzpicture}[x=0.75pt,y=0.75pt,yscale=-1,xscale=1]

\draw    (62,15) -- (62,39) ;
\draw [shift={(62,41)}, rotate = 270] [color={rgb, 255:red, 0; green, 0; blue, 0 }  ][line width=0.75]    (10.93,-3.29) .. controls (6.95,-1.4) and (3.31,-0.3) .. (0,0) .. controls (3.31,0.3) and (6.95,1.4) .. (10.93,3.29)   ;
\draw    (62,74) -- (62,98) ;
\draw [shift={(62,100)}, rotate = 270] [color={rgb, 255:red, 0; green, 0; blue, 0 }  ][line width=0.75]    (10.93,-3.29) .. controls (6.95,-1.4) and (3.31,-0.3) .. (0,0) .. controls (3.31,0.3) and (6.95,1.4) .. (10.93,3.29)   ;
\draw    (62,133) -- (62,157) ;
\draw [shift={(62,159)}, rotate = 270] [color={rgb, 255:red, 0; green, 0; blue, 0 }  ][line width=0.75]    (10.93,-3.29) .. controls (6.95,-1.4) and (3.31,-0.3) .. (0,0) .. controls (3.31,0.3) and (6.95,1.4) .. (10.93,3.29)   ;
\draw   (165,40.2) -- (275,40.2) -- (275,73.33) -- (165,73.33) -- cycle ;
\draw   (166,100.2) -- (277,100.2) -- (277,133.33) -- (166,133.33) -- cycle ;
\draw    (164,233) -- (186.43,250.76) ;
\draw [shift={(188,252)}, rotate = 218.37] [color={rgb, 255:red, 0; green, 0; blue, 0 }  ][line width=0.75]    (10.93,-3.29) .. controls (6.95,-1.4) and (3.31,-0.3) .. (0,0) .. controls (3.31,0.3) and (6.95,1.4) .. (10.93,3.29)   ;
\draw    (131,233) -- (110.48,251.65) ;
\draw [shift={(109,253)}, rotate = 317.73] [color={rgb, 255:red, 0; green, 0; blue, 0 }  ][line width=0.75]    (10.93,-3.29) .. controls (6.95,-1.4) and (3.31,-0.3) .. (0,0) .. controls (3.31,0.3) and (6.95,1.4) .. (10.93,3.29)   ;
\draw   (16,159.2) -- (267,159.2) -- (267,229.33) -- (16,229.33) -- cycle ;
\draw   (7,100.2) -- (117,100.2) -- (117,133.33) -- (7,133.33) -- cycle ;
\draw   (4,41.2) -- (114,41.2) -- (114,74.33) -- (4,74.33) -- cycle ;
\draw    (220,74) -- (220,98) ;
\draw [shift={(220,100)}, rotate = 270] [color={rgb, 255:red, 0; green, 0; blue, 0 }  ][line width=0.75]    (10.93,-3.29) .. controls (6.95,-1.4) and (3.31,-0.3) .. (0,0) .. controls (3.31,0.3) and (6.95,1.4) .. (10.93,3.29)   ;
\draw    (222,133) -- (222,157) ;
\draw [shift={(222,159)}, rotate = 270] [color={rgb, 255:red, 0; green, 0; blue, 0 }  ][line width=0.75]    (10.93,-3.29) .. controls (6.95,-1.4) and (3.31,-0.3) .. (0,0) .. controls (3.31,0.3) and (6.95,1.4) .. (10.93,3.29)   ;
\draw  [dash pattern={on 4.5pt off 4.5pt}] (21,162.2) -- (115,162.2) -- (115,185.33) -- (21,185.33) -- cycle ;
\draw  [dash pattern={on 4.5pt off 4.5pt}] (49,203.33) -- (108,203.33) -- (108,226) -- (49,226) -- cycle ;
\draw  [dash pattern={on 4.5pt off 4.5pt}] (114,203.33) -- (173,203.33) -- (173,226) -- (114,226) -- cycle ;
\draw  [dash pattern={on 4.5pt off 4.5pt}] (178,203.33) -- (237,203.33) -- (237,226) -- (178,226) -- cycle ;
\draw    (220,14) -- (220,38) ;
\draw [shift={(220,40)}, rotate = 270] [color={rgb, 255:red, 0; green, 0; blue, 0 }  ][line width=0.75]    (10.93,-3.29) .. controls (6.95,-1.4) and (3.31,-0.3) .. (0,0) .. controls (3.31,0.3) and (6.95,1.4) .. (10.93,3.29)   ;
\draw   (45,198.33) -- (241,198.33) -- (241,229.17) -- (45,229.17) -- cycle ;

\draw (70,78) node [anchor=north west][inner sep=0.75pt]   [align=left] {OpennessScripter};
\draw (76,258) node [anchor=north west][inner sep=0.75pt]   [align=left] {Satisfied};
\draw (168,259) node [anchor=north west][inner sep=0.75pt]   [align=left] {Violated};
\draw (12,45) node [anchor=north west][inner sep=0.75pt]   [align=left] {TIA Portal - FBD};
\draw (17,107) node [anchor=north west][inner sep=0.75pt]   [align=left] {XML - FBD};
\draw (117,141.2) node [anchor=north west][inner sep=0.75pt]   [align=left] {PLCverif};
\draw (186,46) node [anchor=north west][inner sep=0.75pt]   [align=left] {Specification};
\draw (171,107) node [anchor=north west][inner sep=0.75pt]   [align=left] {Formalized spec.};
\draw (27,167) node [anchor=north west][inner sep=0.75pt]   [align=left] {Instruction List};
\draw (56,207.33) node [anchor=north west][inner sep=0.75pt]   [align=left] {nuXmv};
\draw (122,207.33) node [anchor=north west][inner sep=0.75pt]   [align=left] {CBMC};
\draw (189,207.33) node [anchor=north west][inner sep=0.75pt]   [align=left] {Theta};
\draw (230,78) node [anchor=north west][inner sep=0.75pt]   [align=left] {User};

\end{tikzpicture}

%% file: WEPV042f3.tex
\tikzset{every picture/.style={line width=0.75pt}} 

\begin{tikzpicture}[x=0.75pt,y=0.75pt,yscale=-1,xscale=1]

\draw   (181,40.2) -- (234.5,40.2) -- (234.5,99.43) -- (181,99.43) -- cycle ;
\draw    (126,54.71) -- (179,54.71) ;
\draw [shift={(181,54.71)}, rotate = 180] [color={rgb, 255:red, 0; green, 0; blue, 0 }  ][line width=0.75]    (10.93,-3.29) .. controls (6.95,-1.4) and (3.31,-0.3) .. (0,0) .. controls (3.31,0.3) and (6.95,1.4) .. (10.93,3.29)   ;
\draw    (126,85.71) -- (179,85.71) ;
\draw [shift={(181,85.71)}, rotate = 180] [color={rgb, 255:red, 0; green, 0; blue, 0 }  ][line width=0.75]    (10.93,-3.29) .. controls (6.95,-1.4) and (3.31,-0.3) .. (0,0) .. controls (3.31,0.3) and (6.95,1.4) .. (10.93,3.29)   ;
\draw    (234,69.71) -- (287,69.71) ;
\draw [shift={(289,69.71)}, rotate = 180] [color={rgb, 255:red, 0; green, 0; blue, 0 }  ][line width=0.75]    (10.93,-3.29) .. controls (6.95,-1.4) and (3.31,-0.3) .. (0,0) .. controls (3.31,0.3) and (6.95,1.4) .. (10.93,3.29)   ;
\draw   (181,19) -- (234.5,19) -- (234.5,40) -- (181,40) -- cycle ;
\draw   (107,133) -- (160,133) -- (160,179.43) -- (107,179.43) -- cycle ;
\draw   (107,133) -- (160,133) -- (160,154) -- (107,154) -- cycle ;
\draw  [fill={rgb, 255:red, 0; green, 0; blue, 0 }  ,fill opacity=1 ] (207,108.75) .. controls (207,108.34) and (207.34,108) .. (207.75,108) .. controls (208.16,108) and (208.5,108.34) .. (208.5,108.75) .. controls (208.5,109.16) and (208.16,109.5) .. (207.75,109.5) .. controls (207.34,109.5) and (207,109.16) .. (207,108.75) -- cycle ;
\draw  [fill={rgb, 255:red, 0; green, 0; blue, 0 }  ,fill opacity=1 ] (207,115.75) .. controls (207,115.34) and (207.34,115) .. (207.75,115) .. controls (208.16,115) and (208.5,115.34) .. (208.5,115.75) .. controls (208.5,116.16) and (208.16,116.5) .. (207.75,116.5) .. controls (207.34,116.5) and (207,116.16) .. (207,115.75) -- cycle ;
\draw  [fill={rgb, 255:red, 0; green, 0; blue, 0 }  ,fill opacity=1 ] (207,122.75) .. controls (207,122.34) and (207.34,122) .. (207.75,122) .. controls (208.16,122) and (208.5,122.34) .. (208.5,122.75) .. controls (208.5,123.16) and (208.16,123.5) .. (207.75,123.5) .. controls (207.34,123.5) and (207,123.16) .. (207,122.75) -- cycle ;
\draw   (181,153.2) -- (234.5,153.2) -- (234.5,212.43) -- (181,212.43) -- cycle ;
\draw    (67,167.33) -- (96,167.69) ;
\draw [shift={(98,167.71)}, rotate = 180.7] [color={rgb, 255:red, 0; green, 0; blue, 0 }  ][line width=0.75]    (10.93,-3.29) .. controls (6.95,-1.4) and (3.31,-0.3) .. (0,0) .. controls (3.31,0.3) and (6.95,1.4) .. (10.93,3.29)   ;
\draw    (126,198.71) -- (179,198.71) ;
\draw [shift={(181,198.71)}, rotate = 180] [color={rgb, 255:red, 0; green, 0; blue, 0 }  ][line width=0.75]    (10.93,-3.29) .. controls (6.95,-1.4) and (3.31,-0.3) .. (0,0) .. controls (3.31,0.3) and (6.95,1.4) .. (10.93,3.29)   ;
\draw    (234,182.71) -- (287,182.71) ;
\draw [shift={(289,182.71)}, rotate = 180] [color={rgb, 255:red, 0; green, 0; blue, 0 }  ][line width=0.75]    (10.93,-3.29) .. controls (6.95,-1.4) and (3.31,-0.3) .. (0,0) .. controls (3.31,0.3) and (6.95,1.4) .. (10.93,3.29)   ;
\draw   (181,132) -- (234.5,132) -- (234.5,153) -- (181,153) -- cycle ;
\draw    (164,167.33) -- (178,167.67) ;
\draw [shift={(180,167.71)}, rotate = 181.36] [color={rgb, 255:red, 0; green, 0; blue, 0 }  ][line width=0.75]    (10.93,-3.29) .. controls (6.95,-1.4) and (3.31,-0.3) .. (0,0) .. controls (3.31,0.3) and (6.95,1.4) .. (10.93,3.29)   ;

\draw (135,35) node [anchor=north west][inner sep=0.75pt]   [align=left] {var1};
\draw (134,67) node [anchor=north west][inner sep=0.75pt]   [align=left] {var2};
\draw (240,51) node [anchor=north west][inner sep=0.75pt]   [align=left] {tmp1};
\draw (182,49) node [anchor=north west][inner sep=0.75pt]  [font=\footnotesize] [align=left] {IN1};
\draw (182,80) node [anchor=north west][inner sep=0.75pt]  [font=\footnotesize] [align=left] {IN2};
\draw (207,64) node [anchor=north west][inner sep=0.75pt]  [font=\footnotesize] [align=left] {OUT};
\draw (194,24) node [anchor=north west][inner sep=0.75pt]  [font=\footnotesize] [align=left] {AND};
\draw (108,162) node [anchor=north west][inner sep=0.75pt]  [font=\footnotesize] [align=left] {IN};
\draw (132,162) node [anchor=north west][inner sep=0.75pt]  [font=\footnotesize] [align=left] {OUT};
\draw (121,137) node [anchor=north west][inner sep=0.75pt]  [font=\footnotesize] [align=left] {NOT};
\draw (65,148) node [anchor=north west][inner sep=0.75pt]   [align=left] {tmpn};
\draw (135,184) node [anchor=north west][inner sep=0.75pt]   [align=left] {varn};
\draw (240,164) node [anchor=north west][inner sep=0.75pt]   [align=left] {result};
\draw (182,162) node [anchor=north west][inner sep=0.75pt]  [font=\footnotesize] [align=left] {IN1};
\draw (182,193) node [anchor=north west][inner sep=0.75pt]  [font=\footnotesize] [align=left] {IN2};
\draw (207,177) node [anchor=north west][inner sep=0.75pt]  [font=\footnotesize] [align=left] {OUT};
\draw (198,137) node [anchor=north west][inner sep=0.75pt]  [font=\footnotesize] [align=left] {OR};

\end{tikzpicture}